\documentclass[useAMS, usegraphicx, usenatbib]{mn2e}
\usepackage[ascii]{inputenc}
\usepackage{amsmath}
\usepackage{amsfonts}
\usepackage{amssymb}
\usepackage{multirow,bigdelim}
\usepackage{url}
\usepackage{esdiff}
\newcommand{\subscript}[1]{_{\textrm{#1}}}
\begin{document}
\title[Pop III GRB afterglows]{Gamma-ray burst radio afterglows from Population III stars: Simulation methods and detection prospects with SKA precursors}

\author[Macpherson and Coward]{D. ~Macpherson,$^{1, 2}$ and D. Coward$^1$ \\$^1$UWA School of Physics\\$^2$International Centre for Radio Astronomy Research}

\maketitle
\begin{abstract}
We investigate the prospects of detecting radio afterglows from long Gamma-Ray Bursts (GRBs) from Population III (Pop III) progenitors using the SKA precursor instruments WMA (Murchison Widefield Array) and ASKAP (Australian SKA Pathfinder). We derive a realistic model of GRB afterglows that encompasses the widest range of plausible physical parameters and observation angles. We define the best case scenario of Pop III GRB energy and redshift distributions. Using probability distribution functions fitted to the observed microphysical parameters of long GRBs, we simulate a large number of Pop III GRB afterglows to find the global probability of detection. We find that ASKAP may be able to detect 35\% of Pop III GRB afterglows in the optimistic case, and 27\% in the pessimistic case. A negligible number will be detectable by MWA in either case. Detections per image for ASKAP, found by incorporating intrinsic rates with detectable timescales, are as high as $\sim$ 6000 and as low as $\sim$ 11, which shows the optimistic case is unrealistic. We track how the afterglow flux density changes over various time intervals and find that, because of their very slow variability, the cadence for blind searches of these afterglows should be as long as possible. We also find Pop III GRBs at high redshift have radio afterglow lightcurves that are indistinguishable from those of regular long GRBs in the more local universe.
\end{abstract}


\section{Introduction}
Gamma-Ray Bursts (GRBs) are the brightest explosions in the universe. These transient events outshine entire galaxies for a short time. The degree of variability and short life of GRBs points towards a compact source. Typical isotropic equivalent energies lie in the region of $10^{53}$erg. Such high energy from a compact object can be explained if the emission is narrowly collimated and highly relativistic \citep{rhoads1997,sari1999,rhoads1999,frail2001,bloom2003,chandra2012}. The most plausible hypothesis for the production of a GRB is a system consisting of a stellar mass black hole surrounded by an accretion torus within a magnetic field. Angular momentum in the accretion torus is extracted via the Blandford-Znajek mechanism \citep{bz77} to produce a polar outflow of highly relativistic plasma. Velocity variations in the outflow lead to internal shocks where the electrons are accelerated to produce gamma-ray photons.

In addition, the outflow produces external shocks when it interacts with an ambient medium, leading to synchrotron radiation (afterglow of the GRB). The peak frequency of this afterglow decreases as the outflow loses energy; typically what begins as a predominantly X-ray afterglow, over time becomes mostly optical, infrared, and eventually a radio afterglow \citep{meszaros1997, sari1998, piran2004, zhang2007, gao2013}. A similar mechanism explains the relativistic outflow and radio lobes seen in active galactic nuclei (AGN), with the difference being mainly one of scale: AGN involve supermassive black holes, are less bright, less variable, and much longer lived \citep{mirabel2004, ghisellini2005, nemmen2012, bromberg2011}.

GRBs are classified as either short/hard or long/soft (as well as more exotic types such as under-luminous and ultra-long). Within the framework of GRBs being produced by black hole formation/accretion, the distinction between the two categories can be explained by fundamentally different methods of formation or accretion onto a stellar mass black hole. Short duration GRBs with hard gamma-ray spectra are thought to originate from a merger of compact objects, two neutron stars or  a black hole and a neutron star \citep{goodman1986,woosley1993,ruffert1999,bloom2002,aloy2005,rezzolla2011}. Long duration GRBs with soft gamma-ray spectra are thought to originate from the core collapse of a relatively massive star \citep{woosley1993,macfadyen1999,frail2001,bloom2002}.

Short GRBs are unsuited as probes of Pop III stars, as the production of the compact stellar remnant progenitors introduces a large degree of separation between the Pop III star and the short GRB. Where one progenitor is a black hole, the only attribute connected to the original star is its mass. Add to this the time taken to produce the compact remnants, but mostly in inspiral time for the two remnants to merge, short GRBs are too far removed from the original stars to be suitable probes of Pop III stars \citep{bloom2002}. Furthermore, short GRBs tend to have very faint afterglows.

On the other hand, the rapid evolution and short life of a massive star which culminates in a long GRB links those GRBs closely with the initial star formation. Therefore long GRBs from Pop III stars can be used to constrain the formation history of the first generation of stars.

The potential of utilising GRBs as tools to study Pop III stars has been strengthened by observations of very high-z afterglows (eg GRB090429B z $\sim$ 9.4 \citep{cucchiara2011}). The formation history of Pop III stars has been theoretically studied by multiple authors. Such work includes the ionisation rate of the IGM and universal chemical evolution history that explain star formation rates in the local universe. Theoretical Pop III star formation rates (SFRs) peak between redshifts 6 and 20 \citep{tumlinson2006,trenti2009,maio2010,desouza2011,johnson2012,wise2012}. Others build models to calculate the rate at which stars form out of the primordial environment, with the earliest limits of Pop III star formation being placed at z $\sim$ 60 \citep{naoz2007}. In the most optimistic case the peak of formation is at z $\sim$ 6, with appreciable formation still occurring at z $\sim$ 3. With GRBs detectable at z $\sim$ 9.4, it should be possible to use GRBs to probe what is possibly the peak of Pop III star formation, or at least to constrain the redshift limits at which Pop III formation ceased.

As a foundation for our simulation, we employ previous studies to derive a model that is capable of accepting the broad distribution of physical parameters that determine the afterglow lightcurve. We refine the method of accounting for spherical shape and beaming of the afterglow emission region to improve the efficiency of the calculations, and expanded it to allow for arbitrary orientation angles. Our refinements result in a new model that can take practically any set of plausible physical parameters of a GRB, at any orientation angle and observing frequency, and return the lightcurve over $10^{12}$ seconds within a minute on a desktop computer.

We have adopted this approach (see Macpherson et al. 2013) in favour of the construction of a luminosity function for Pop III GRB afterglows for three reasons. Firstly, given that Pop III GRBs are still highly speculative objects, constructing a realistic luminosity function is ambiguous. Secondly, the luminosity function considers the brightness at a specific time, usually one day after the burst. Most radio afterglows are still increasing in brightness at this time \citep{chandra2012}, thus using a luminosity function, one would underestimate the proportion of detectable afterglows. Thirdly, the suitability of luminosity functions to transients like GRB afterglows is dubious; they are more suited to objects whose luminosity can be considered constant or stable for most intents and purposes.

The paper is organised as follows: the afterglow simulation method is described in Section 2, and Section 3 describes the distributions of GRB parameters used in the model. The results of the simulations of Pop III GRB afterglows are presented and discussed in Section 4. Section 5 summarises our main results, the detectability of Pop III afterglows by SKA precursors.

\section{Afterglow Simulation Method}
As a foundation to simulate the light curves of the GRB afterglows we use a method based primarily on Mesler et al. (2014). We begin by calculating the initial parameters of the ejecta: the bulk Lorentz factor ($\Gamma_0$), the ejecta mass ($M_{\rm ej}$), and initial radius $r_0$.

The formulation we use to calculate the initial bulk Lorentz factor $\Gamma_0$ is taken from \citet{toma2011}, with an initial time $t_0$ set to 1 second,

\begin{align}
\Gamma_0 & = \left(\dfrac{E_{\rm iso}}{4\pi nm_pc^5t_0^3}\right)^{1/8},
\end{align}
where $E_{\rm iso}$ is the isotropic equivalent energy of the GRB, $m_p$ is the proton mass, and $n$ is the external medium density.

$M_{\rm ej}$, and the deceleration radius $r_0$ at which $\Gamma$ begins to evolve, are found using the method described in \citet{panaitescu2000}:

\begin{align}
M_{\rm ej} & = \dfrac{E}{c^2\Gamma}, \label{eq_Mej}\\
r_0 & = \left(\dfrac{(3-s)E}{4\pi Am_pc^2\Gamma^2}\right)^{1/3-s}. \label{eq_r0}
\end{align}

In the above equations, $E$ is the total energy of the fireball, which is found by scaling $E_{\rm iso}$ by the fraction of a hemisphere projected by the initial jet opening angle $\theta_j$, that is $E = E_{\rm iso}(1-\cos \theta_j)$. The values A and s in equation \ref{eq_r0} are properties of the external medium density, i.e. $n(r) = Ar^{-s}$. In the case of an isotropic medium, $s = 0$ and $A = n$.

The mass $m_0$ swept up by the fireball by radius $r_0$ is found by \citep{huang1999,panaitescu2000,pe'er2012}:

\begin{align}
m(r_0) & = \dfrac{M_{ej}}{\Gamma}.
\end{align}

The evolution of $\Gamma$ is evaluated numerically over intervals of $r$, for which the rate of change of $\Gamma$ is:

\begin{align}
\label{dGamma}
\diff{\Gamma}{m} &= - \dfrac{\hat{\gamma}(\Gamma^2 - 1) - (\hat{\gamma} - 1)\Gamma\beta^2}{M_{\rm ej} + \epsilon m + (1-\epsilon)m(2\hat{\gamma}\Gamma - (\hat{\gamma} - 1)(1+\Gamma^{-2}))}, \\
\hat{\gamma} &= \dfrac{4\Gamma + 1}{3\Gamma}, \\
\epsilon &= \epsilon_e \dfrac{t_{\rm syn}^{\prime-1}}{t_{\rm syn}^{\prime-1} + t_{\rm exp}^{\prime-1}}, \\
t_{\rm syn}' &= \dfrac{6\pi m_e c}{\sigma_T B^{\prime 2} \gamma_m}, \\
B^{\prime 2} &= 32 \pi \epsilon_B n m_p c^2 \Gamma(\Gamma - 1),
\end{align}
where $\hat{\gamma}$ is the adiabatic index of the jet material, $\epsilon$ is the radiative efficiency, $\epsilon_e$ is the fraction of blast wave energy in the electrons, $t_{\rm syn}^{\prime}$ is the comoving synchrotron cooling timescale \citep{dai1999,rhoads1999}, $t_{\rm exp}^{\prime}$ is the comoving age of the jet, $\sigma_T$ is the Thompson cross section, $B^{\prime}$ is the comoving magnetic field \citep{mesler2014}, and $\epsilon_B$ is the magnetic field energy fraction of the blast wave.

The minimum Lorentz factor of injected electrons, $\gamma_m$, is calculated in different ways depending on the value of the electron energy index, p. Where p$>$2,

\begin{equation}
\gamma_m = 1 + \epsilon_e(\Gamma - 1)\dfrac{m_p(p-2)}{m_e(p-1)}, \label{steep}
\end{equation}

\citep{johannesson2006,mesler2014}. If the electron energy distribution is unusually flat (p$<$2),

\begin{align}
\gamma_m &=	1 + \left(\epsilon_e(\Gamma - 1)\gamma_u^{p-2}\dfrac{m_p(2-p)}{m_e(p-1)}\right)^{1/(p-1)} , \label{flat} \\
\gamma_u &\simeq \left(\dfrac{3e_c}{\sigma_T B'}\right)^{1/2} \label{max_gamma}.
\end{align}

Equation \ref{flat} \citep{bhatta2001,dai2001} depends upon the maximum injected electron Lorentz factor $\gamma_u$, here approximated by Equation \ref{max_gamma} \citep{toma2011}, where $e_c$ is the electron charge.

The differential mass swept up d$m$ is approximated by:

\begin{equation}
\diff{m}{r} = 4\pi r^2 n(r)m_p
\end{equation}

\citep{huang1999,pe'er2012}. With these equations we find $\Gamma$ and $m$ in terms of $r$, which \citet{mesler2014} relate to local rest frame time $t$ as:

\begin{equation}
\diff{t}{r} = \dfrac{1}{c\beta\Gamma(\Gamma + \sqrt{\Gamma^2 - 1})}.
\end{equation}

The evolving opening half angle $\theta_j$ is found by calculating $c_s$, the speed of sound within the blast wave (Equation \ref{csound}, \citet{mesler2014}). Over a time interval d$t$, the blast wave spreads sideways by an amount $c_s$d$t$, in addition to expansion due to radial motion. Defining $a$ as the lateral size of the blast wave, we calculate $\theta_j$ after a small change in radius d$r$ by:

\begin{align}
c_s &= c\Gamma\sqrt{\dfrac{\hat{\gamma}(\hat{\gamma} - 1)(\Gamma - 1)}{1 + \hat{\gamma}(\Gamma - 1)}}, \label{csound}\\
a &= r\tan(\theta_j) + (c_s{\rm d}t), \\
\theta_j &= \arctan(a/r).
\end{align}

The synchrotron emission spectrum of the blast wave depends on three characteristic Lorentz factors: the minimum injection Lorentz factor $\gamma_m$, the cooling Lorentz factor $\gamma_c$, and the absorption Lorentz factor $\gamma_a$.

The Lorentz factor of cooling electrons is

\begin{equation}
\gamma_c  = \dfrac{6\pi m_ec}{B'^2\sigma_Tt'(Y+1)}
\end{equation}

\citep{sari1998, bhatta2001,johannesson2006,toma2011,mesler2014}, where $Y$ is the Compton parameter, whose value depends on the relative values of the characteristic Lorentz factors as shown below.

\begin{align}
Y = \left\lbrace 
\begin{array}{ll}
	\gamma_m\gamma_c\tau_e &  \gamma_m > \gamma_c \\
	\tau_e\gamma_m^{p-1}\gamma_c^{(3-p)} &  \gamma_c > \gamma_m \\
	\tau_e(C_2^{2-p}\gamma_c^7\gamma_m^{7(p-1)})^{1/(p+5)} &  \gamma_a > \textrm{max}(\gamma_m, \gamma_c),
\end{array}
\right.
\end{align}
where $\tau_e$ is the optical depth:
\begin{equation}
\tau_e = \dfrac{\sigma_Tm(r)}{4\pi m_pr^{\prime 2}},
\end{equation}
and
\begin{equation}
C_2 = \dfrac{5e_c\tau_e}{\sigma_TB'}.
\end{equation}

Because the equations of the Compton parameter $Y$ depend on the relative values of $\gamma_c$, $\gamma_m$, and $\gamma_a$, one must find the solution to $\gamma_c$ which is self-consistent. This is complicated further by the fact that $\gamma_a$ also depends on the relative values of the other characteristic Lorentz factors, as described below:

\begin{align}
\gamma_a = \left\lbrace
	\begin{array}{ll}
	C_2^{0.3}\gamma_c^{-1/2} &  \gamma_a < \gamma_c < \gamma_m \\
	C_2^{0.3}\gamma_m^{-1/2} &  \gamma_a < \gamma_m < \gamma_c \\
	(C_2\gamma_c)^{1/6} &  \gamma_c < \gamma_a < \gamma_m \\
	(C_2\gamma_m^{p-1})^{1/(p+4)} &  \gamma_m < \gamma_a < \gamma_c \\
	(C_2\gamma_c\gamma_m^{p-1})^{1/(p+5)} &  \gamma_a > \textrm{max}(\gamma_m, \gamma_c),
	\end{array}
\right.
\end{align}

We approach this problem by finding the values of $\gamma_c$ in all possible scenarios, then select whichever one is self-consistent. For instance, if the equation for $Y$ in the fast cooling regime (i.e. $\gamma_m > \gamma_c$) results in $\gamma_c$ greater than $\gamma_m$, obviously this solution is invalid and the jet is not in the fast cooling regime. Then, if the $Y$ equation for the slow cooling regime also yields $\gamma_c > \gamma_m$, the jet is definitely in the slow cooling regime and the value of $\gamma_c$ is valid.

Similarly in finding the $\gamma_a$, we find all possible values, under every arrangement of the characteristic Lorentz factors, and find that which is self-consistent.

These characteristic Lorentz factors transform to characteristic frequencies via:

\begin{equation}
\nu(\gamma) = \dfrac{3\Gamma e_cB'}{(1-\beta)4\pi m_ec},
\end{equation}
where $\beta$ is the velocity (in terms of $c$) corresponding to the Lorentz factor $\gamma$.

Then the peak flux density is calculated via

\begin{equation}
F_{\nu}^{\textrm{max}} = \dfrac{(1+z)\sqrt{3}\phi_p e_c^3}{4\pi d_L^2m_em_pc^2} \beta_m^2 \Gamma B' m(r),
\end{equation}
where $\phi_p$ is the p-dependent factor defined in \citet{wijers1999}, $d_L$ is the luminosity distance to the GRB, and $\beta_m$ is the velocity of the lowest-energy injected electrons. Prior to the jet break at time $t_j$, $F_{\nu}^{\textrm{max}}$ remains approximately constant. After the jet break, the peak flux density evolves with time as $F_{\nu}^{\textrm{max}} = F_{\nu, t_j}^{\textrm{max}} (t_j/t)$ \citep{rhoads1999,bhatta2001}, where $F_{\nu, t_j}^{\textrm{max}}$ is the peak flux density calculated at the point of the jet break.

The jet break marks the stage at which the centre of the jet head becomes causally linked with its outer edge; when the bulk Lorentz factor of the jet is less than the inverse of its angular separation ($\Gamma < 1/\theta_j$). In the model we use, $\theta_j$ evolves with all other quantities. This evolution is slow at early times, accelerating at the point of jet break.

The peak flux density is emitted at the median of the three characteristic frequencies. The slope of the spectrum between characteristic frequencies depends upon those frequencies' relative values, as detailed below (all frequencies in the frame of the local ISM).

In the fast cooling ($\nu_c<\nu_m$) regime:

\begin{align}
F_{\nu} = F_{\nu}^{\textrm{max}} \left\lbrace
	\begin{array}{lr}
	(\nu/\nu_a)^2(\nu_a/\nu_c)^{1/3} & \nu<\nu_a<\nu_c<\nu_m \\
	(\nu/\nu_c)^{1/3} & \nu_a<\nu<\nu_c<\nu_m \\
	(\nu_c/\nu)^{1/2} & \nu_a<\nu_c<\nu<\nu_m \\
	(\nu/\nu_m)^{-p/2}(\nu_c/\nu_m)^{1/2} & \nu_a<\nu_c<\nu_m<\nu \\
	(\nu/\nu_c)^2(\nu_c/\nu_a)^{5/2} & \nu<\nu_c<\nu_a<\nu_m \\
	(\nu/\nu_a)^{5/2} & \nu_c<\nu<\nu_a<\nu_m \\
	(\nu_a/\nu)^{1/2} & \nu_c<\nu_a<\nu<\nu_m \\
	(\nu/\nu_m)^{-p/2}(\nu_a/\nu_m)^{1/2} & \nu_c<\nu_a<\nu_m<\nu \\
	(\nu/\nu_c)^2(\nu_c/\nu_a)^{5/2} & \nu<\nu_c<\nu_m<\nu_a \\
	(\nu/\nu_a)^{5/2} & \nu_c<\nu<\nu_a \\
	(\nu/\nu_a)^{-p/2} & \nu_c<\nu_m<\nu_a<\nu \\
	\end{array}
\right.
\end{align}

In the slow cooling ($\nu_m<\nu_c$) regime:

\begin{align}
F_{\nu} = F_{\nu}^{\textrm{max}} \left\lbrace
	\begin{array}{lr}
	(\nu/\nu_a)^2(\nu_a/\nu_m)^{1/3} & \nu<\nu_a<\nu_m<\nu_c \\
	(\nu/\nu_m)^{1/3} & \nu_a<\nu<\nu_m<\nu_c \\
	(\nu/\nu_m)^{-(p-1)/2} & \nu_a<\nu_m<\nu<\nu_c \\
	(\nu/\nu_c)^{-p/2}(\nu_c/\nu_m)^{-(p-1)/2} & \nu_a<\nu_m<\nu_c<\nu \\
	(\nu/\nu_m)^2(\nu_m/\nu_a)^{5/2} & \nu<\nu_m<\nu_a<\nu_c \\
	(\nu/\nu_a)^{5/2} & \nu_m<\nu<\nu_a<\nu_c \\
	(\nu/\nu_a)^{(p-1)/2} & \nu_m<\nu_a<\nu<\nu_c \\
	(\nu/\nu_c)^{p/2}(\nu_c/\nu_a)^{(p-1)/2} & \nu_m<\nu_a<\nu_c<\nu \\
	(\nu/\nu_m)^2(\nu_m/\nu_a)^{5/2} & \nu<\nu_m<\nu_c<\nu_a \\
	(\nu/\nu_a)^{5/2} & \nu_m<\nu<\nu_a \\
	(\nu/\nu_a)^{-p/2} & \nu_m<\nu_c<\nu_a<\nu \\
	\end{array}
\right.
\end{align}

Hence the flux density is obtained at any given frequency $\nu$ for a local observer. However, we have not yet accounted for the spherical nature of the emitting region or its relativistic velocity. The spherical shape introduces a slight delay to radiation emitted from angles further from the observer's line of sight. Were the flow not relativistic, this delay could be calculated from simple geometric arguments; effects of special relativity complicate this angle dependent delay. Further, the Doppler beaming factor for high $\Gamma$ amplifies emissions along the line of sight while reducing emissions at angles greater than 1/$\Gamma$.

From the appendix of \citet{mesler2014} the equal arrival time surface is

\begin{equation}
\label{eqarrive}
t = \textrm{constant} = \int\dfrac{(1-\beta\cos\theta)}{c\beta}dr,
\end{equation}
where it is suggested that this equation be used to find the emission radius, $r$, of radiation from a section of the jet at an angle $\theta$, as measured from the explosion centre, to the line of sight, by integrating over an increasing $r$ with constant $\theta$ until reaching the observation time $t$. This must be done for $0<\theta<\theta_j$ to find the equal time arrival surface $r$ as a function of $\theta$.

Repeated integration of equation (\ref{eqarrive}) to find $r$ for incremental values of $\theta$ is a time consuming process. It is more efficient to do the opposite: to find $\theta$ for incremental values of $r$. The equation can be rearranged into $t = - r\cos\theta/c + \int {\rm d}r/c\beta$, and from there it can be shown that

\begin{equation}
\theta = \arccos\left(\begin{array}{c}\displaystyle
\dfrac{r - \displaystyle\int_{r_i}^{r} \frac{{\rm d}r}{\beta}}{r_i}\end{array}\right).
\end{equation}

Thus radiation emitted at some smaller radius $r_i = r - \Delta r$, reaching the observer simultaneously with that emitted at radius $r$, will have been at an angle $\theta$ to the line of sight.

The total observed flux density is given by

\begin{equation}
\label{eq:obsfd}
F_{\nu} = \iint\limits_{\Omega_j} \dfrac{L_{\nu'}^{\prime}[r(\theta)]\mathcal{D}^3}{\Omega_j} {\rm d}\cos\theta {\rm d}\phi,
\end{equation}
where $L_{\nu'}^{\prime}[r(\theta)]$ is the total radiated luminosity at the comoving frequency $\nu'$ at $r(\theta)$, $L_{\nu'}^{\prime} = F_{\nu} / \Gamma$, $\mathcal{D}$ is the Doppler factor $\mathcal{D} = 1/\Gamma(1-\beta\cos\theta)$, and $\Omega_j$ is the jet solid angle.

When the observer is oriented precisely on the jet axis, $\int {\rm d}\phi = 2\pi$, as the emitting region appears a full circle. In more general terms, if the observer is offset from the jet axis by some angle $\theta_i$, $\int {\rm d}\phi = 2\pi$ only for angles $\theta \leq \theta_j - \theta_i$, and evaluates to 0 when $\theta<\theta_i-\theta_j$ and $\theta>\theta_i+\theta_j$. When $\theta_i-\theta_j < \theta < \theta_i+\theta_j$, $\int {\rm d}\phi = 2\arccos((\theta^2 + \theta_i^2 - \theta_j^2)/2\theta\theta_i)$.

Thus we iteratively reduce $r$ by an appropriately small fraction $\Delta r$, for each value calculating the emission angle $\theta$. As long as $\theta<\theta_i+\theta_j$, we also calculate the total luminosity $L_{\nu'}^{\prime}(r) = F_{\nu}(r) / \Gamma(r)$, Doppler factor $\mathcal{D}(r, \theta) = 1/\Gamma(r)(1-\beta(r)\cos\theta)$, and $\int {\rm d}\phi$. Once done, we integrate $L_{\nu'}^{\prime}[r(\theta)]\mathcal{D}^3$ from 0 to $\theta_i+\theta_j$, and divide by $\Omega_j$ to find the total observed flux density.

\begin{figure}
\centering
\includegraphics[width=90mm]{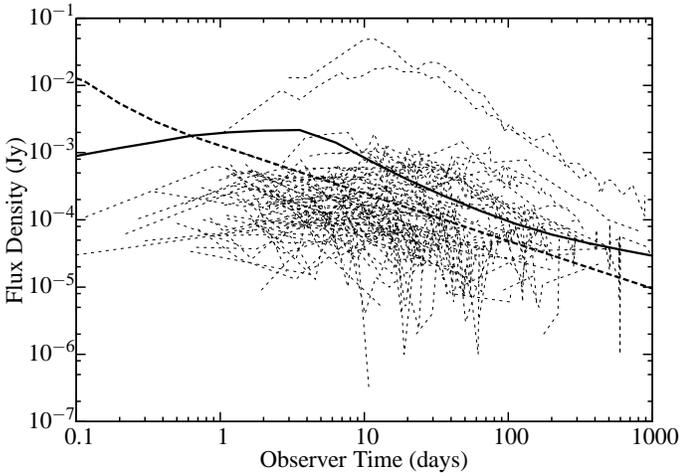}
\caption{\label{radio_comp}Comparison of observed (dotted lines) and simulated fiducial GRB 8.5GHz radio afterglow lightcurves in the case of an isotropic external medium (solid line) and a wind medium (dashed line).}
\end{figure}

Figure \ref{radio_comp} shows how the model's afterglow lightcurves compare with observations in both a constant density medium and a wind medium. The common parameters of the simulated GRBs are: $E_{\rm iso}$ = 10$^{53}$ erg, $z$ = 1, $\theta_j$ = 0.1 rad, $p$ = 2.3, $\epsilon_e$ = 0.1, $\epsilon_B$ = 0.01. In the case of the isotropic medium, the density $n$ is 1cm$^{-3}$. In the case of the wind medium, the density is inversely proportional to $r^{2}$, and the values of the deceleration radius $r_0$ and density at that radius $n(r_0)$ are set to the fiducial values in Panaitescu \& Kumar (2000, equations (8) and (9)) for a GRB from a Wolf-Rayet star. These are plotted over a sample of observed GRB radio afterglow lightcurves at frequencies at or near 8.5GHz. The data for these radio afterglows was provided by Poonam Chandra (2014, private communication).

The simulated lightcurves in Figure \ref{radio_comp} are comparable to observations, and this shows that the simulation is realistic. An $E_{\rm iso}$ of $10^{53}$ ergs and redshift of 1 applied to both cases (constant density and wind medium) are roughly median values as measured through observations, and produces roughly median lightcurves. We do not consider whether this $E_{\rm iso}$ and $z$ are true medians of the intrinsic distributions, as both the lightcurves and the apparent parameters are drawn from observations, hence are subject to the same selection effects.

The differences between the wind density and constant density medium cases are small. The lightcurve in the wind density case has a steeper rise to peak earlier than in the constant density case. At later times, and for the majority of the time when the afterglows are observable, the lightcurves tend to follow each other with less than half an order of magnitude between their flux densities.

We will only consider constant density media in the remainder of this paper. To properly construct a wind density profile requires a robust estimate of the mass loss rate of the progenitor star. In the case of Pop III stars, this is far from trivial. It has been argued that the rate of radiation-driven mass loss from a Pop III star should be minimal due to the lack of absorption lines in its mostly hydrogen envelope and environment. On the other hand, it is argued that in order for a massive Pop III star to collapse while retaining enough angular momentum to form a GRB, it must undergo chemically homogeneous evolution \citep{yoon2012}, which would dredge heavier elements produced in the core up to the outer layers, restoring radiation driven mass loss. Furthermore, chemically homogeneous evolution requires the star to have a high rotation velocity, which implies that the kinematic mass loss should be high. This last point may or may not be influential to GRB afterglows, as the kinematics that drive mass loss are greatest at the star's equator, and zero at the poles where the relativistic GRB jets are expected to launch.

Finding a satisfactory solution to the problem of the properties of a wind medium around a Pop III star would be a separate and substantial investigation. Given the relatively small difference in the resultant lightcurves as shown in Figure \ref{radio_comp}, we keep to the simpler case of assuming constant density media.

\section{Distributions of GRB physical parameters}
The relevant physical parameters that determine GRB afterglows are the total isotropic equivalent energy $E_{\rm iso}$, redshift z, the jet half-opening angle $\theta_j$, density of the surrounding medium $n$, the electron energy distribution index $p$, the electron energy fraction $\epsilon_e$, and the magnetic field energy fraction $\epsilon_B$.

We have no reason to expect that the distributions of the parameters $\theta_j$, $n$, $p$, $\epsilon_e$, or $\epsilon_B$ for Pop III progenitor GRBs will differ from those of normal GRBs. Many GRB simulations use a set of `fiducial' parameters as a way of estimating typical GRB lightcurves. In this work we investigate GRBs where the parameters are allowed to vary as observations show, thus creating a realistic set of GRB lightcurves. The fiducial parameters provide a starting point and reality check for the parameter distributions. A typical set of fiducial parameters from \citet{toma2011} is $\theta_j$ = 0.1, $n$ = 1.0 cm$^{-3}$, $p$ = 2.3, $\epsilon_e$ = 0.1, and $\epsilon_B$ = 0.01.

We do expect that the distributions of $E_{\rm iso}$ and $z$ are closely linked to the distribution of progenitor mass and the SFR, respectively. As these distributions are uncertain, we investigate two extreme scenarios for the $E_{\rm iso}$ and $z$ distributions: under the most optimistic and most pessimistic combinations of Pop III initial mass function (IMF) and SFR.

\subsection{GRB isotropic equivalent energy $E_{\rm iso}$}
The energy reservoir of a long GRB is associated with the mass of its progenitor star. The amount of energy released is proportional to the mass of the accretion torus around the nascent black hole, which in turn is proportional to the total mass of the star \citep{toma2011}. \citet{nakauchi2012} and \citet{suwa2011} simulated the production of GRBs from collapsing Pop III stars of various initial masses, and found the resultant $E_{\rm iso}$s. The results were summarised in Macpherson et al. (2013; Figure 1), from which a linear relation between progenitor mass and resultant GRB $E_{\rm iso}$ can be seen, described as 

\begin{equation}
E\subscript{iso}(\times 10^{55} {\rm erg}) = 0.0124(M/{\rm M}_{\odot})+0.6507.
\end{equation}

This allows us to relate the energy distribution of GRBs to the IMF of the progenitor stars. 

However, we do not expect every Pop III star to produce a GRB. According to the work of \citet{yoon2012}, there is a certain window in mass and rotation speed in which GRB formation is possible. The progenitor must have an initial mass between $\sim$ 12 - 84 M$_{\odot}$, with a rotation velocity close to the break-up limit. From figure 12 of \citet{yoon2012} and assuming a mass-independent distribution of Pop III stellar rotation speed, we generate a function of the probability of a Pop III star being able to produce a GRB as a function of mass. We find that the probability of a Pop III star forming a GRB is highest for a mass of $\sim 60$M$_{\odot}$. 

The IMF of Pop III stars is unknown. Previous studies provide several possible alternatives to the standard Salpeter power law. These include lognormal distributions \citep{tumlinson2006} and high mean Gaussian distributions \citep{desouza2011}. Of those we found it was the Gaussian distribution with $\mu = 55$M$_{\odot}$ and $\sigma = 15$M$_{\odot}$ of \citet{scannapieco2003} as used in \citet{desouza2011} that was the most top heavy and, combined with the \citet{yoon2012} mass and rotation criteria, provided the greatest overall probability of Pop III GRB formation.

\begin{figure}
\centering
\includegraphics{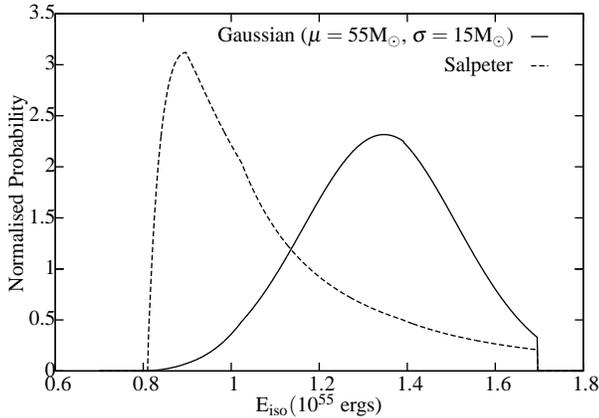}
\caption{\label{E_dist}Normalised $E\subscript{iso}$ distributions of Population III GRBs under two scenarios of the Pop III IMF; a high mean gaussian \citep{scannapieco2003} (solid line) and the Salpeter power law (dashed line).}
\end{figure}

Assuming a mass-independent stellar rotation distribution, we derive from figure 12 of \citet{yoon2012} mass-dependent rotation limits for GRB formation. Combining this with the IMF and using the progenitor mass-$E_{\rm iso}$ relation, we generate the Pop III GRB $E_{\rm iso}$ probability function. A graphical representation of the normalised Pop III GRB $E_{\rm iso}$ distributions is shown in Figure \ref{E_dist}.

\subsection{Redshift distribution}
Progenitors of long GRBs are massive stars, whose lives are relatively short. The occurrence of long GRBs is therefore closely associated with the SFR. We can use theoretical estimates of the Pop III SFR to find the redshift distribution of Pop III GRBs. Based on the approach described in \citet{bromm2002}, the non-normalised GRB redshift probability function is: 

\begin{eqnarray}
\label{pgrb}
P_{\rm GRB}(z) = \dfrac{\Psi_*(z)d_L^2}{(1+z)^3\sqrt{\Omega_m(1+z)^3 + \Omega_{\Lambda}}} {\rm d}z,
\end{eqnarray}
where $\Psi_*(z)$ is the Pop III SFR, and cosmological parameters $\Omega_m$ and $\Omega_{\Lambda}$ are 0.267 and 0.734 respectively.

\begin{figure}
\includegraphics{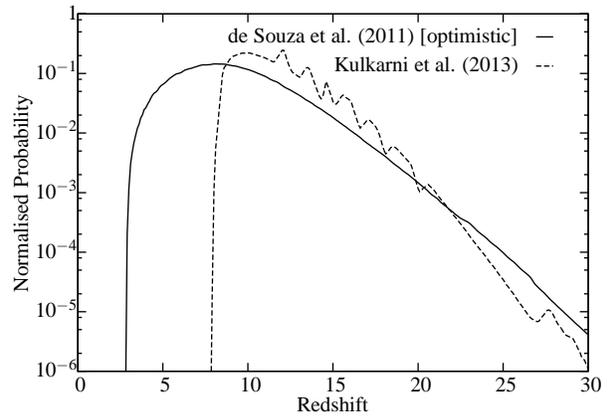}
\caption{Pop III GRB redshift probability density functions under the most optimistic (solid line) and the most pessimistic (dashed line) Pop III SFRs.}
\label{GRB_dist}
\end{figure}

The highest Pop III SFR found in the literature is the most optimistic case of \citet{desouza2011}, while the lowest is that of \citet{kulkarni2013}. The normalised Pop III GRB redshift distributions based on these SFRs are shown in Figure \ref{GRB_dist}.

\subsection{Initial jet opening angle $\theta_j$}
Analysis of the observed opening angles of GRBs has suggested that $\theta_j$ evolves, where angles are narrower at high redshift \citet{yonetoku2005}. However, there is no clear physical reason for this evolution, and it is alternatively possible that the apparent evolution is a selection effect \citep{lu2012}. GRBs with narrower opening angles have a higher apparent brightness, and are therefore visible from greater distances. At a high redshift, an observer is less likely to see a GRB with a larger opening angle, so the population of GRBs at that redshift would appear dominated by those with smaller $\theta_j$. \citet{lu2012} found using a simulated set of GRBs that the apparent redshift dependent $\theta_j$ distribution could be reproduced if the intrinsic $\theta_j$ distribution were lognormal, with median e$^{\mu}$ = 0.0537 and standard deviation $\sigma$ = 0.6.

This median $\theta_j$ is slightly lower than the fiducial value, but not by an egregious amount. We therefore adopt this distribution in our simulations.

\subsection{Circumburst density n}
From \citet{panaitescu2002,chevalier2004,piro2005,cenko2011} we take circumburst density measurements of twelve GRBs, shown in Table \ref{parametertable}.

\begin{table}
\caption{Observationally derived microphysical parameters of GRBs}
\label{parametertable}
\begin{tabular}{lrrrr}
\hline
GRB	 & 	Density (cm$^{-3}$)	& p & $\epsilon_e$ & $\epsilon_B$ \\
\hline																	
970228 &  & 2.44$^{8}$ &  & \\
970508 & 0.75$^{2}$ & 2.29$^{2,3,8}$ & 0.19$^{1,2,3}$ & 0.1685$^{1,2,3,6}$ \\
971214 & & 2.2$^{8}$ & & \\
980329 & & 2.69$^{3,8}$ & 0.12$^{3}$ & 0.17$^{3}$ \\
980519 & 0.14$^{2}$ & 2.87$^{2,8}$ & 0.11$^{2}$ & 0.000035$^{2}$ \\
980703 & & 2.64$^{3,8}$ & 0.27$^{3}$ & 0.0018$^{3}$ \\
990123 & 0.0019$^{4}$ & 2.135$^{2,8}$ & 0.13$^{2}$ & 0.00074$^{2}$ \\
990510 & 0.29$^{4}$ & 1.945$^{2,8}$ & 0.025$^{2}$ & 0.0052$^{2}$ \\
991208 & 18$^{2}$ & 1.53$^{2}$ & 0.056$^{2}$ & 0.035$^{2}$ \\
991216 & 4.7$^{2}$ & 1.36$^{2}$ & 0.014$^{2}$ & 0.018$^{2}$ \\
000301C	& 27$^{4}$ & 1.43$^{2}$ & 0.062$^{2}$ & 0.072$^{2}$ \\
000418 & 27$^{2}$ & 2.04$^{2}$ & 0.076$^{2}$ & 0.0066$^{2}$ \\
000926 & 22$^{4}$ & 2.58$^{2,3,8}$ & 0.125$^{2,3}$ & 0.0435$^{2,3}$ \\
010222 & 1.7$^{2}$ & 1.695$^{2,8}$ & 0.43$^{2}$ & 0.000067$^{2}$ \\
011121 & & 2.5$^{5}$ & & 0.5$^{5}$ \\
011211 & 3$^{5}$ & 2.4$^{5}$ & 0.0025$^{5}$ & 0.01$^{5}$ \\
050801 & & 2.64$^{9}$ & & \\
050802 & & 2.62$^{9}$ & & \\
050904 & & & 0.0309$^{7}$ & \\
051109A & & 2.08$^{9}$ & & \\
060124 & & 2.02$^{9}$ & & \\
060729 & & 2.22$^{9}$ & & \\
061121 & & 1.88$^{9}$ & & \\
090323 & & & 0.07$^{10}$ & 0.0089$^{10}$ \\
090328 & & 2.26$^{10}$ & 0.11$^{10}$ & 0.0019$^{10}$ \\
090902B & 0.00056$^{10}$ & 2.21$^{10}$ & 0.13$^{10}$ & 0.33$^{10}$ \\
090926A	& & 2.13$^{10}$ & 0.33$^{10}$ & 0.081$^{10}$ \\
\hline
\end{tabular}
\\
{\scriptsize 1: \cite{wijers1999} \par
2: \cite{panaitescu2002} \par
3: \cite{yost2003} \par
4: \cite{chevalier2004} \par
5: \cite{piro2005} \par
6: \cite{vanderhorst2005} \par
7: \cite{gou2007} \par
8: \cite{starling2008} \par
9: \cite{curran2009} \par
10: \cite{cenko2011} \par}
Where multiple references are given, the value shown is the numerical average of the reference values.
\end{table}

Fitting to a lognormal distribution\footnote{ Python 2.7.3, scipy 0.14.0 \\scipy.stats.lognorm.fit(data, floc = 0)} yields distribution parameters e$^{\mu}$ = 0.88 and $\sigma$ = 3.5. The Kolmogorov-Smirnov (K-S) test of the data to this distribution\footnote{ scipy.stats.kstest(data, `lognorm', args = (3.5, 0, 0.88))} shows a 90\% probability that the data came from this distribution. The distribution's median value of 0.88 is not dissimilar to the fiducial density of 1.0 cm$^{-3}$.

\subsection{Electron energy index p}
Previous studies have derived a larger number of p values, as well as possible distributions of this parameter. Several theoretical works concluded that p should have a near-universal value of $\sim$ 2.2 - 2.3, with very little variation \citep{achterberg2001,lemoine2003}. Observationally, p varies much more widely, as seen in Table \ref{parametertable}.

A Gaussian distribution that fits the data in Table \ref{parametertable} has $\mu$ = 2.2, $\sigma$ = 0.4. The K-S test of this data against a distribution with these parameters\footnote{scipy.stats.kstest(data, `norm', args = (2.2, 0.4))} yields a 99\% probability that this data came from such a distribution. Further, the mean of 2.2 is consistent with the fiducial value of 2.3.

\subsection{Electron ($\epsilon_e$) and magnetic ($\epsilon_B$) energy fractions}
Observationally derived values for $\epsilon_e$ are shown in Table \ref{parametertable}. At first glance, the values roughly conform to a slight variance around the fiducial value of 0.1.

Using the K-S test, we find a 74\% probability that the observed values of $\epsilon_e$ are from a lognormal distribution with e$^{\mu}$ = 0.09 and $\sigma$ = 1.2. The median value compares well with the fiducial $\epsilon_e$ of 0.1.

Using the K-S test, we find a 98\% probability that the $\epsilon_B$values shown in Table \ref{parametertable} are from a lognormal distribution with e$^{\mu}$ = 0.012, $\sigma$ = 2.6. We are again assured by the proximity of the median to the fiducial $\epsilon_B$ of 0.01.

\section{Results}
With the parameter distributions described in Section 3, we use the GRB afterglow model described in Section 2 to simulate GRBs from Pop III stellar collapse. The probability distribution functions of the parameters are used to generate pseudo-random values, which are then fed into the simulation to produce lightcurves.

\begin{figure}
\includegraphics{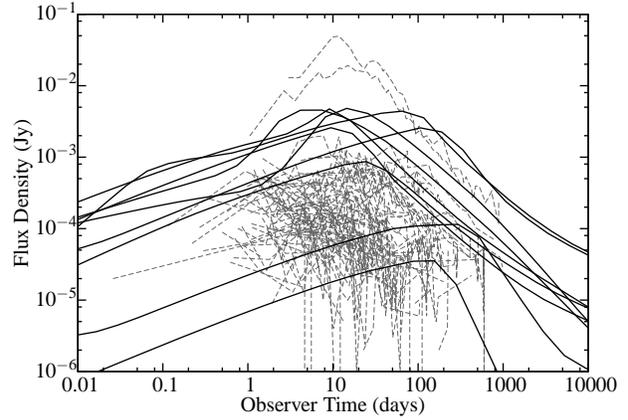}
\caption{Comparison of observed GRB radio afterglows (dotted lines) \citep{chandra2012} and simulated Pop III GRB afterglows (solid lines) at 8.5 GHz.}
\label{chan_p3}
\end{figure}

Firstly, by testing those simulations where the observation angle is within the initial opening angle (the GRB is aligned with the observer), and comparing to observed afterglows, we determine that the afterglows of aligned Pop III GRBs will be observable. Using the 8.5 GHz radio afterglow data provided by Poonam Chandra (2014, private communication), the comparison of simulated Pop III GRB afterglows to observed GRB afterglows is shown in Figure \ref{chan_p3}. The figure shows that the simulated lightcurves are similar to observations, so the afterglows of Pop III GRBs should easily be detectable at frequencies near 8.5 GHz with existing instruments.

As shown in Figure \ref{chan_p3}, there is nothing in the shape of the afterglow radio lightcurves to distinguish between GRB progenitors. The lightcurves from our simulations preferentially occupy the higher peak flux density region of the spread of observed lightcurves, but no individual burst would be noticeably different. One may have expected that the higher energy budget presumed available to Pop III GRBs would produce brighter afterglows, and indeed in the local ISM frame this would be the case. However, the high redshift of these GRBs then reduces the observed flux densities to values typical of local GRBs.

The simulations show that the radio afterglows of Pop III GRBs should not appear very different from normal GRBs. Conversely, it also means that an observed radio afterglow could potentially be a high redshift Pop III GRB.

\begin{figure}
\includegraphics{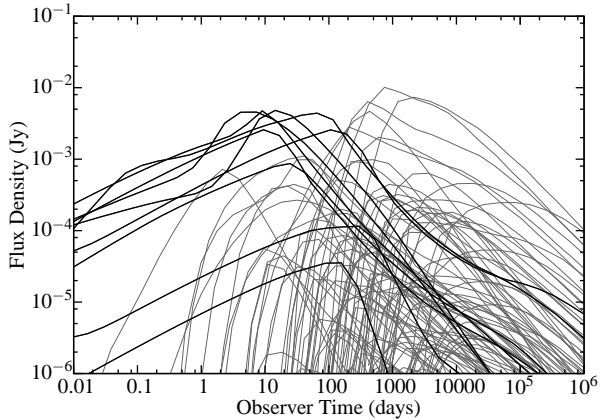}
\caption{8.5 GHz radio afterglows of simulated Pop III GRBs when the observer is on-axis (black lines) and off-axis (grey lines).}
\label{all_8.5}
\end{figure}

Having shown that Pop III GRB afterglows are detectable, in theory, we also consider the contribution of unaligned `orphan' GRBs, where the observing angle is initially outside the jet opening angle. Figure \ref{all_8.5} shows that the flux densities can reach detectable levels after the jet break .

\subsection{Detection by SKA precursors}
In this study we consider two Square Kilometre Array (SKA) precursor instruments; the Murchison Widefield Array (MWA) and the Australian SKA Pathfinder (ASKAP).

MWA is a low frequency wide field phased array instrument, operating in the frequency range 80-300 MHz. The defined characteristic performance of this instrument is given at the central frequency of 150 MHz with a 30 MHz bandwidth, where it has a nominal field of view of 610 deg$^2$ and angular resolution of between $\sim$ 2 $\sim$ 3 arcmin \citep{tingay2013}. The sky noise at this frequency, combined with the angular resolution, results in MWA being largely confusion limited. The 5$\sigma$ detection limit is 6 mJy, which the instrument achieves in a matter of minutes. When simulating the MWA lightcurves we integrate over a band 30 MHz wide centred on 150 MHz, using 6 mJy as the sensitivity limit.

ASKAP's operational frequency range is 0.7-1.8 GHz, with a bandwidth of 300 MHz. With the instrument configured for an angular resolution of 18", it has a 5$\sigma$ point source sensitivity of 29 $\mu$Jy in a 1 hour observation \citep{johnston2009}. In our simulations of the ASKAP lightcurves we integrate over a band 300 MHz wide centred on 1.25 GHz, and use 29 $\mu$Jy as the sensitivity limit.

\begin{figure}
\includegraphics{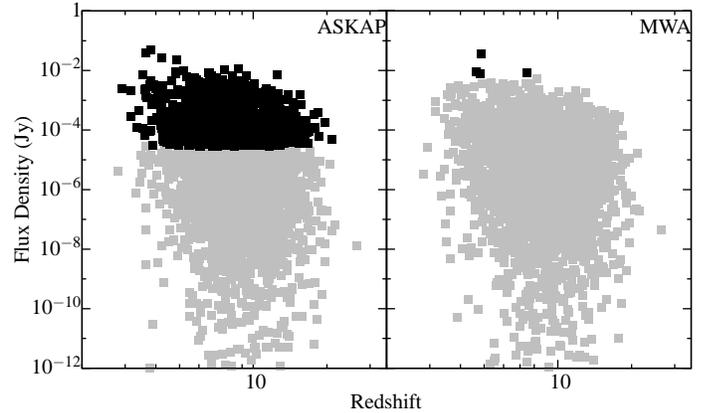}
\caption{Peak flux densities of simulated lightcurves of Pop III GRB radio afterglows in the frequency bands of ASKAP and MWA plotted against redshift in the most optimistic case. Black points show afterglows exceeding the detection threshold, while grey points are undetectable afterglows.}
\label{peaks_opt}
\end{figure}

Figure \ref{peaks_opt} shows the peak flux densities of the simulated afterglows plotted against redshift for ASKAP and MWA. Grey points indicate that the peak flux density is less than the detection limit, and black points are above the detection limit.

Figure \ref{peaks_opt} shows that ASKAP will detect many more Pop III GRB afterglows than MWA. 34.8\% of all afterglows could be detectable by ASKAP. Only 4 of the three thousand simulated afterglows achieved a peak flux density greater than 6mJy at 150MHz, a detectability rate of 0.13\%. Closer analysis of the data reveals 89.4\% of aligned afterglows will exceed ASKAP's detection limit, and that the majority (70\%) of those will do so within one day after the burst.

The simulated lightcurves whose peak flux are shown in Figure \ref{peaks_opt} assume a very optimistic scenario for Pop III SFR and IMF. Properly investigating the possibility of Pop III GRB radio afterglow detection requires that we also consider the worst case (and likely more realistic) scenario. The most pessimistic derived Pop III SFR is that of \citet{kulkarni2013}, where the redshift of peak Pop III star formation occurs at higher $z$, mostly at $z\gtrsim 10$, and never at $z < 7.5$.

\begin{figure}
\includegraphics{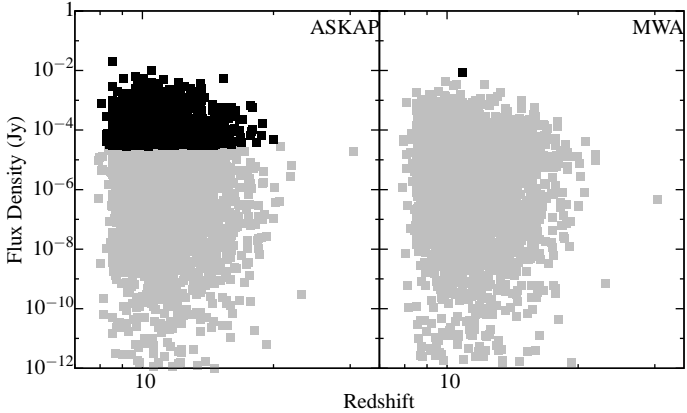}
\caption{Peak flux densities of simulated lightcurves of Pop III GRB radio afterglows in the bands observable by ASKAP and MWA plotted against redshift in the most pessimistic scenario of Pop III SFR and IMF. Black points show afterglows exceeding the detection threshold, while grey points show undetectable afterglows.}
\label{peaks_pess}
\end{figure}

The Kulkarni SFR also assumes the least top-heavy, thus for us the most pessimistic, Pop III IMF of a Salpeter power law distribution with a low mass cut-off of 1M$_{\odot}$. The pessimistic SFR and IMF yield new probability distribution functions for redshift and $E_{\rm iso}$. The simulated afterglows' peak flux densities with these PDFs are shown in Figure \ref{peaks_pess}.

The figure shows that even in the pessimistic scenario of Pop III formation, the resultant GRB radio afterglows will be detectable by ASKAP. There is a reduction in the proportion of detectable afterglows; in this scenario, 82.7\% of aligned afterglows and 26.9\% of all afterglows exceed the limit of ASKAP. Only a single simulation (0.03\% of all afterglows) exceeded the detection limit for MWA.

\subsection{Pop III GRB afterglow detection rates}
Our simulations give a representation of the detectability of Pop III GRBs. To assess the probability of making a detection, we must estimate the intrinsic rate at which these GRBs occurs. In constructing our sets of simulated GRB afterglows, we have used theoretical SFRs and IMFs, both of which affect the formation rate of GRBs. Using the approach of Bromm \& Loeb (2002), the GRB rate above a certain redshift is as follows:

\begin{equation}
\label{eq:grbrate}
N_{\rm GRB}(>z) = \dfrac{4\pi c\Delta t_{\rm obs}}{H_0} \int_{z}^{\infty}\dfrac{\eta_{\rm GRB}\Psi_*(z)d_L^2}{(1+z)^3\sqrt{\Omega_m(1+z)^3 + \Omega_{\Lambda}}} {\rm d}z,
\end{equation}
where $\Psi_*$ is the Pop III SFR, $\Delta t_{\rm obs}$ is the observation time interval, set to 1 yr to cancel the yr$^{-1}$ component of the SFR, $H_0$ is the Hubble constant, and $\eta_{\rm GRB}$ is the GRB formation efficiency:

\begin{equation}
\eta_{\rm GRB} = f_{\rm GRB} \dfrac{\int_{M_L}^{M_U} \phi(m)r(m) {\rm d}m}{\int_{M_L}^{M_U} m\phi(m) {\rm d}m},
\end{equation}
where M$_U$ and M$_L$ are the upper and lower limits of $\phi(m)$, the IMF. $r(m)$ is the probability of satisfying the \citet{yoon2012} rotation speed criteria as a function of mass. $f_{\rm GRB}$ is the GRB formation fraction, being the fraction of potential progenitors which actually produce GRBs, set at 0.001 \citep{langer2006}.

In the most optimistic case of Pop III GRB formation, we use the most optimistic Pop III SFR of \citet{desouza2011}, combined with a Gaussian IMF with $\mu=55$M$_{\odot}$ and $\sigma=15$M$_{\odot}$ \citep{scannapieco2003} as used in \citet{desouza2011}. With this IMF, $\eta_{\rm GRB}$ is $6.35\times 10^{-6}$M$_{\odot}^{-1}$. Applying this and the optimistic SFR to equation \ref{eq:grbrate} for all redshifts results in an intrinsic Pop III GRB rate of $4.55\times 10^3$yr$^{-1}$. We can scale this rate by the fraction of the sky covered by the ASKAP, and find that Pop III GRBs would occur at a rate of 3.31 yr$^{-1}$ within its 30 sq. deg. FOV.

Multiplying the average rate of GRBs per day per FOV by the average number of days that the afterglows are detectable, results in the average number of detections per image. Where this number is greater than 1, the implication is that, at any given time, there should be multiple simultaneously observable GRB afterglows. Where this number is less than 1, it implies multiple observations will be required per detection.

We only perform this analysis for ASKAP, for as shown in Figures \ref{peaks_opt} and \ref{peaks_pess}, the probability of MWA detecting a Pop III GRB afterglow is negligible.

\begin{figure}
\includegraphics{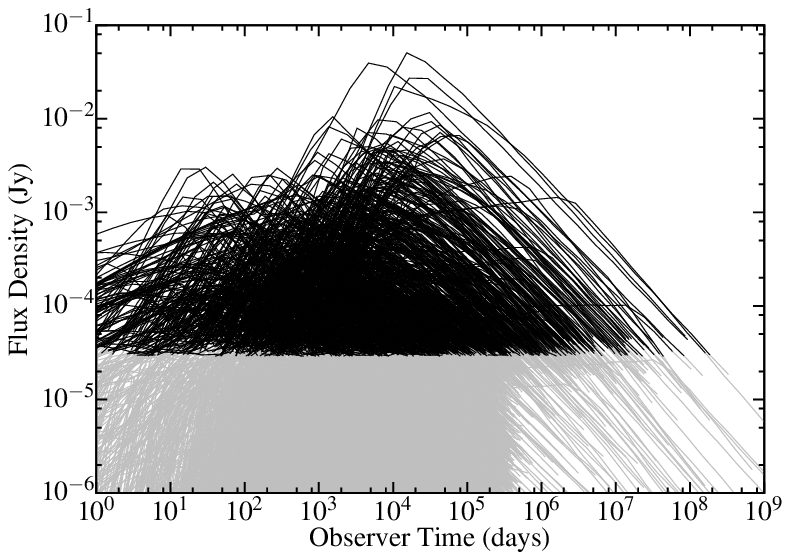}
\includegraphics{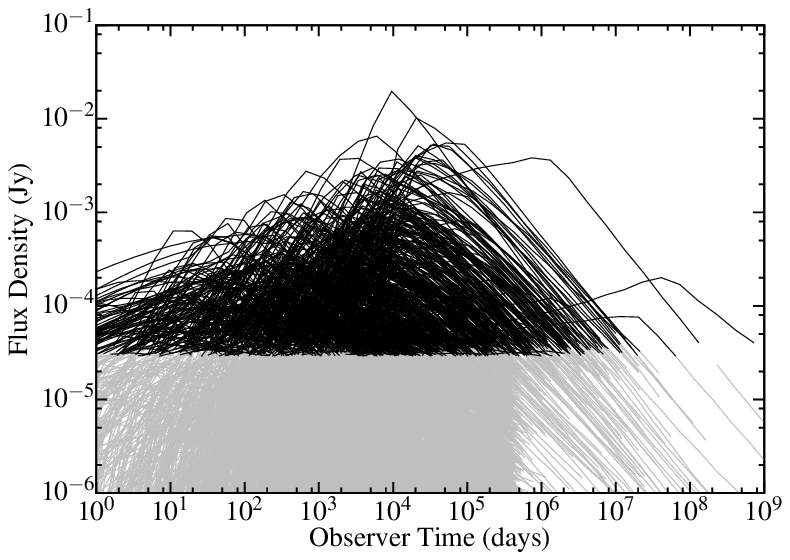}
\caption{Simulated lightcurves of Pop III GRB radio afterglows at a frequency band 1.10-1.40 GHz, observable by ASKAP. Grey segments of the lightcurves are below ASKAP detection threshold of 29$\mu$Jy; black segments are above this threshold. {\bf Top} The optimistic scenario. {\bf Bottom} The pessimistic scenario.}
\label{askap}
\end{figure}

Figure \ref{askap} (Top) shows the simulated GRB afterglows resulting from using this SFR and IMF. Analysing the data, we find that the average length of time that a Pop III GRB afterglow remains detectable by ASKAP is 6.60$\times 10^5$ days, inclusive of those which never exceed the detection threshold. If we exclude the invisible GRBs, then the remainder are detectable by ASKAP for an average of 1.90$\times 10^6$ days.

In this optimistic case, for ASKAP we have 9.06$\times 10^{-3}$ Pop III GRBs per day multiplied by 6.60$\times 10^5$ days, resulting in a detection number of 5.98$\times 10^3$. These $\sim 6000$ Pop III GRB afterglows in every 30 sq. deg. image taken by ASKAP would, on average, be separated by 4.8 arcmin. This scenario should therefore be easily proven or disproven by ASKAP.

The most pessimistic case of Pop III GRB formation rate consists of the Pop III SFR derived by \citet{kulkarni2013}, adopting a Salpeter power law IMF in the range 1-100M$_{\odot}$. This IMF combined with the aforementioned formation criteria results in $\eta_{\rm GRB} = 1.82\times 10^{-6}$M$_{\odot}^{-1}$. Applying this efficiency and the pessimistic SFR to equation \ref{eq:grbrate} results in an intrinsic Pop III GRB rate for all redshifts of 7.95 yr$^{-1}$. This equates to a rate of 1.58$\times 10^{-5}$yr$^{-1}$ within ASKAP's FOV.

From Figure \ref{askap} (Bottom), the average time an afterglow can be detected by ASKAP is 6.72$\times 10^5$ days. The average number of detectable Pop III GRB afterglows per ASKAP image is 10.6, with an average angular separation between afterglows of 1.2 degrees.

Since the proportions of afterglows detectable by MWA is negligible in both pessimistic and optimistic scenarios, we do not calculate detection rates for this instrument.

\subsection{Detection of prompt gamma-ray emission}
So far we have not been concerned with the orientation of the GRBs, treating the small subset of directed bursts right along with those undirected bursts which only become visible at late times. In terms of performing a blind search, whether or not the observer was within the initial GRB opening angle is irrelevant. We now look at the prospects of a triggered follow up of Pop III GRBs from the prompt gamma-ray detection by {\it Swift}.

\begin{figure}
\includegraphics{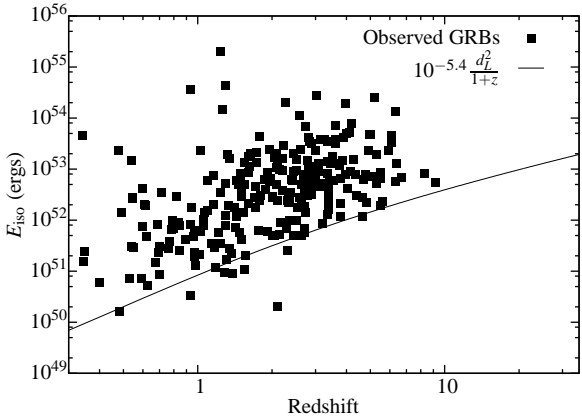}
\caption{Observed GRBs $E_{\rm iso}$ as a function of redshift, roughly following the expected $\frac{d_L^2}{1+z}$ relation.}
\label{lower_lim}
\end{figure}

The first question is whether or not the prompt gamma-ray emission from a Pop III GRB will be sufficiently luminous to trigger a GRB satellite. We assume that peak gamma-ray luminosity scales with $E_{\rm iso}$ and we observe that the minimum detectable peak luminosity, and therefore minimum detectable $E_{\rm iso}$, increases with redshift as $\dfrac{d_L^2}{1+z}$. Figure \ref{lower_lim} plots the limiting $E_{\rm iso}$ with redshift following the expected $E_{\rm iso,lim}\propto\dfrac{d_L^2}{1+z}$. To ensure a consistent sample of $E_{\rm iso}$ data, we use the Butler online catalogue \emph{Swift} BAT Integrated Spectral Parameters\footnote{\url{http://butler.lab.asu.edu/Swift/bat_spec_table.html}}. This catalogue, an extension of \citet{butler2007,butler2010}, circumvents the nominal BAT upper energy of 150 keV to produce values of $E_{\rm iso}$ through a Bayesian approach.

The figure shows that at z = 35, which is the upper limit for the theoretical Pop III SFRs adopted in this paper, the minimum observable $E_{\rm iso}$ is less than 10$^{54}$ergs. It is less than the minimum $E_{\rm iso}$ of Pop III GRBs assumed in this paper (see Figure \ref{E_dist}), so in this scenario the prompt emissions will be sufficiently luminous to trigger {\it Swift}.

The second question is the rate at which both {\it Swift} is within the initial opening angle of a Pop III GRB, and that GRB is within the FOV of the satellite. Within {\it Swift}'s 2 str FOV, Pop III GRBs occur at a rate of 724 yr$^{-1}$ in the optimistic case, and 1.27 yr$^{-1}$ in the pessimistic case. Through our simulations, the observer was within the initial opening angle of approximately 4\% of all GRBs. In the optimistic case {\it Swift} would detect 29 Pop III GRBs yr$^{-1}$, and in the pessimistic case 0.05 yr$^{-1}$ (one Pop III GRB every 20 years).

Following {\it Swift} detection, ASKAP would be capable of detecting $\sim$90\% of the 29 GRBs yr$^{-1}$ (26 yr$^{-1}$), in the optimistic case, and have an $\sim$83\% chance of detecting a 1/20 year Pop III GRB in the pessimistic case. MWA should not be considered to follow up Pop III GRB prompts in any case, as our simulations have shown that a) the chances of achieving sufficient flux density for MWA to detect are extremely low, and b) the sufficient flux density occurs a very long time after the burst.

\subsection{Optimal observing cadence for detection}
The optimal cadence to search for slow transients such as GRB radio afterglows are those that have the highest average flux variance between images. We consider an observation plan lasting ten years, and populate a number of theoretical FOVs with simulated afterglow lightcurves. We took a range of observation intervals, from observing every day for the full ten years, to having only two observations separated by ten years. Over this range we find the average changes in afterglow flux density over a single time interval.

For the optimistic case with ASKAP, we find that the average change in flux density increases with the length of the interval, up to the maximum interval of 10 years, where the average change in flux density was $\sim 6\times 10^{-5}$Jy. The pattern continued in the pessimistic case, where the maximum average flux change of $\sim 2.7 \times 10^{-5}$ Jy occurs at the maximum interval of 10 years.

We find the optimum cadence for these searches with instruments like ASKAP will be greater than ten years. For practical purposes, the results indicate that when it comes to a blind search for Pop III GRB radio afterglows, one should look at the longest feasible cadence.

\section{Conclusions}
We find the radio afterglows of Pop III GRBs are detectable by ASKAP, via the application of a realistic afterglow model, under the following assumptions: 1: Pop III stars formed at high redshift, with higher masses and therefore shorter lives than Pop I/II stars. 2: Pop III stars could produce GRBs in the manner of stellar core collapse, providing certain criteria of mass and rotation were met. 3: The resulting GRBs would be more energetic than those from Pop I/II stellar collapse. 4: The microphysical parameters affecting GRBs and their afterglows do not change with stellar population.

Our simulations show that the radio afterglow lightcurves of high redshift Pop III GRBs will be indistinguishable from those of low redshift Pop I/II GRBs. This undermines the primary motivation for searching for Pop III GRBs: the study of Pop III star formation history. The only indication that a radio afterglow might belong to a Pop III star would be if the GRB were at high redshift ($z > 5$). The fact that the radio emission is similar to that of a low redshift GRB ($z < 5$) would indicate Pop III-type energy. The distinction cannot be based solely on the radio afterglow lightcurves, as unfortunately there is no way to use radio afterglow data to determine redshift.

Even if radio observations could identify high redshift and orphan GRB afterglows, those observations would not by themselves provide evidence that the progenitor were Pop I/II or III. Observations at higher frequencies will be necessary, but at these redshifts the Gunn-Peterson trough of Lyman-$\alpha$ absorption rules out frequencies higher than near-infrared. With advanced space infrared telescopes (e.g. James Webb Space Telescope) it may be possible to obtain spectra of high redshift GRB afterglows, which would reveal the composition of their environments. However, given that optical/infrared afterglows tend to decay quickly, it is likely that such observations would only be possible for GRBs directed at and triggering a GRB satellite.

It is expected that the peak gamma-ray luminosity of Pop III GRBs will be high enough to trigger a GRB satellite, such as {\it Swift}, despite their higher redshifts. The rate of prompt detection relies on the alignment rate scaled by the sky fraction covered by GRB satellites. {\it Swift} could detect as many as 29 Pop III GRB prompts yr$^{-1}$ in the most optimistic case, and as few as one every 20 years in the pessimistic case.

Simulations of the afterglow lightcurves of Pop III GRBs in the optimistic case show that 89\% of the aligned bursts will be detectable by ASKAP, 70\% of which within 1 day. ASKAP's sensitivity lends it to timely follow-up of GRB prompts, although rapid follow-up interrupting its normal observation schedule should not be necessary due to the longevity of the afterglows. Observations made some tens of days after the GRB may identify an afterglow with a Pop III progenitor if it has a brightness typical of a low redshift Pop I/II progenitor and is shown to be high redshift (z $>$ 5).

Our simulations do not find any afterglows of aligned Pop III GRBs detectable to MWA. The total number of detections was negligible (0.13\%), and only possible when near their peak flux density, which is extremely late at MWA frequencies. MWA is unsuited for follow up of Pop III GRB afterglows, or indeed GRBs in general.

Altogether 35\% of all aligned and unaligned Pop III GRB afterglows will be detectable by ASKAP in the optimistic case. Since, on average, only 4\% of all GRBs will be aligned with the observer, the feasibility of detecting the unaligned `orphan' Pop III GRB afterglows with these instruments is important. The average number of detections per image is the product of the intrinsic rate of Pop III GRBs within the FOV with the average time an afterglow is detectable. For ASKAP this is 5.98$\times 10^3$ detectable afterglows per image, a number which is unrealistically high. The number of detectable afterglows predicted in this case is high enough to be relatively easily disproved, and we expect ASKAP transient searches to dismiss this upper limit.

In the pessimistic case, analysis of the simulations shows ASKAP capable of detecting 83\% of aligned afterglows and 27\% of all afterglows, with an average of 10.6 detections per image. MWA detections are negligible (0.03\%).

The average change in afterglow flux density over a time interval increases with the length of that interval, at least up to the maximum tested interval of ten years. This average flux variance seems to be of the order of the flux density limit of detection. In performing a blind or archival search for orphan afterglows, one should use the longest feasible cadences and search for flux decay.

While detecting orphan radio afterglows may be possible, our results show that it will not be possible to distinguish progenitors from orphan radio afterglow data alone. Combining such data with other detections may, in principle, reveal new insight into progenitors. For instance, detections of an orphan afterglow in multiple frequencies from sub-mm to radio, and a non-detection in optical/infrared frequencies may be due to redshifted Lyman-$\alpha$ absorption indicative of a high redshift and high energy GRB progenitor. The key question would be the proportion of Pop III GRBs whose orphan optical/infrared afterglows would be detectable if not for redshifted Lyman-$\alpha$ absorption. This question we will investigate in a follow-up study.

\section{Acknowledgements}
We wish to acknowledge the assistance of Dr. Poonam Chandra in supplying radio afterglow data. D.M. Coward is supported by an Australian Research Council Future Fellowship (FT100100345).  We thank the reviewer for providing suggestions that have improved the clarity of our results.


\begin{thebibliography}{widestlabel}
\bibitem[\protect\citeauthoryear{Achterberg et al.}{2001}]{achterberg2001}Achterberg A., Gallant Y., Kirk J., Guthmann, A., 2001, MNRAS, 328, 393
\bibitem[\protect\citeauthoryear{Aloy et al.}{2005}]{aloy2005}Aloy M.A., Janka H.-Th., Muller E., 2005, A\&A, 436, 273
\bibitem[\protect\citeauthoryear{Bhattacharya}{2001}]{bhatta2001}Bhattacharya D., 2001, BASI, 29, 107
\bibitem[\protect\citeauthoryear{Blandford \& Znajek}{1977}]{bz77}Blandford R.D., Znajek R.L., 1977, MNRAS, 179, 433
\bibitem[\protect\citeauthoryear{Bloom et al.}{2002}]{bloom2002}Bloom J.S., Kulkarni S.R., Djorgovski S.G., 2002, ApJ, 123, 1111
\bibitem[\protect\citeauthoryear{Bloom et al.}{2003}]{bloom2003}Bloom J.S., Frail D.A., Kulkarni S.R., 2003, ApJ, 594, 674
\bibitem[\protect\citeauthoryear{Bromberg et al.}{2011}]{bromberg2011}Bromberg O., Nakar E., Piran T., Sari R., 2011, ApJ, 740, 100
\bibitem[\protect\citeauthoryear{Bromm \& Loeb}{2002}]{bromm2002}Bromm V. and Loeb A., 2002, ApJ, 575, 111
\bibitem[\protect\citeauthoryear{Butler at al.}{2007}]{butler2007}Butler N.R., Kocevski D., Bloom J.S., Curtis J.L., 2007, ApJ, 671, 656
\bibitem[\protect\citeauthoryear{Butler et al.}{2010}]{butler2010}Butler N.R., Bloom J.S., Poznanski D., 2010, ApJ, 711, 495
\bibitem[\protect\citeauthoryear{Cenko et al.}{2011}]{cenko2011}Cenko S. et al., 2011, ApJ, 732, 29
\bibitem[\protect\citeauthoryear{Chandra \& Frail}{2012}]{chandra2012}Chandra P., Frail D.A., 2012, ApJ, 746, 156
\bibitem[\protect\citeauthoryear{Chevalier \& Li}{2004}]{chevalier2004}Chevalier R., Li Z.-Y., 2004, ApJ, 606, 369
\bibitem[\protect\citeauthoryear{Cucchiara}{2011}]{cucchiara2011}Cucchiara A. et al., 2011, ApJ, 736, 7
\bibitem[\protect\citeauthoryear{Curran et al.}{2009}]{curran2009}Curran P., Starling R., van der Horst A., Wijers R., 2009, MNRAS, 395, 580
\bibitem[\protect\citeauthoryear{Dai et al.}{1999}]{dai1999}Dai Z.G., Huang Y.F., Lu T., 1999, ApJ, 520, 634
\bibitem[\protect\citeauthoryear{Dai \& Cheng}{2001}]{dai2001}Dai Z.G., Cheng K.S., 2001, ApJ, 558, L109
\bibitem[\protect\citeauthoryear{de Souza et al.}{2011}]{desouza2011}de Souza R., Yoshida N., Ioka K., 2011, A\&A, 533, A32
\bibitem[\protect\citeauthoryear{Frail et al.}{2001}]{frail2001}Frail D.A. et al., 2001, ApJ, 562, L55
\bibitem[\protect\citeauthoryear{Gao et al.}{2013}]{gao2013}Gao H., Lei W.-H., Zou Y.-C., Wu X.-F., Zhang B., 2013, NewAR, 57, 141
\bibitem[\protect\citeauthoryear{Ghisellini}{2005}]{ghisellini2005}Ghisellini G., 2005, IJMPA, 20, 6991
\bibitem[\protect\citeauthoryear{Goodman}{1986}]{goodman1986}Goodman J., 1986, ApJ, 308, L47
\bibitem[\protect\citeauthoryear{Gou et al.}{2007}]{gou2007}Gou L.-J., Fox D., M\'{e}sz\'{a}ros P., 2007, ApJ, 668, 1083
\bibitem[\protect\citeauthoryear{Huang et al.}{1999}]{huang1999}Huang Y. F., Dai Z. G., Lu T., 1999, MNRAS, 309, 513
\bibitem[\protect\citeauthoryear{Johannesson et al.}{2006}]{johannesson2006}Johannesson G., Bjornsson G., Gudmundsson E.H., 2006, ApJ, 647, 1238
\bibitem[\protect\citeauthoryear{Johnson et al.}{2012}]{johnson2012}Johnson J., Dalla V., Kochfar S.,2012, MNRAS, 428, 1857
\bibitem[\protect\citeauthoryear{Johnston et al.}{2009}]{johnston2009}Johnston S., Feain I., Gupta N., 2009, in Saikia D., Green D., Gupta Y., Venturi T., eds, ASPC Series Vol. 407, Science with the Australian Square Kilometre Array Pathfinder (ASKAP). ASPC
\bibitem[\protect\citeauthoryear{Kulkarni et al.}{2013}]{kulkarni2013}Kulkarni G., Rollinde E., Hennawi J., Vangioni E., 2013, ApJ, 772, 93
\bibitem[\protect\citeauthoryear{Langer \& Norman}{2006}]{langer2006}Langer N., Norman C.A., 2006, ApJ, 638, L63
\bibitem[\protect\citeauthoryear{Lemoine \& Pelletier}{2003}]{lemoine2003}Lemoine M., Pelletier G., 2003, ApJ, 589, 73L
\bibitem[\protect\citeauthoryear{Lu et al.}{2012}]{lu2012}Lu R.-J., Wei J.-J., Qin S.-F., Liang E.-W., 2012, ApJ, 745, 168
\bibitem[\protect\citeauthoryear{MacFadyen \& Woosley}{1999}]{macfadyen1999}MacFadyen A.I., Woosley S.E., 1999, ApJ, 524, 262
\bibitem[\protect\citeauthoryear{Macpherson et al.}{2013}]{macpherson2013}Macpherson D., Coward D., Zadnik M., 2013, ApJ, 779, 73
\bibitem[\protect\citeauthoryear{Maio et al.}{2010}]{maio2010}Maio U., Ciardi B., Dolag K., Tornatore L., Kochfar S., 2010, MNRAS, 407, 1003
\bibitem[\protect\citeauthoryear{Mesler et al.}{2014}]{mesler2014}Mesler R.A., Whalen D.J., Smidt J., Fryer C.L., Lloyd-Ronning N.M., Pihlstrom Y.M., 2014, ApJ, 787, 91
\bibitem[\protect\citeauthoryear{M\'{e}sz\'{a}ros \& Rees}{1997}]{meszaros1997}M\'{e}sz\'{a}ros P., Rees M.J., 1997, ApJ, 476, 232
\bibitem[\protect\citeauthoryear{Mirabel}{2004}]{mirabel2004}Mirabel I.F., 2004, in Schonfelder V., Lichti G., Winkler C., eds, ESA SP-552, 5th \textit{INTEGRAL} Workshop on the \textit{INTEGRAL} Universe. ESA Publications Division, Noordwijk, p. 175 
\bibitem[\protect\citeauthoryear{Nakauchi et al.}{2012}]{nakauchi2012}Nakauchi D., Suwa Y., Sakamoto T., Kashiyama K., Nakamura T., 2012, ApJ, 759, 128
\bibitem[\protect\citeauthoryear{Naoz \& Bromberg}{2007}]{naoz2007}Naoz S., Bromberg O., 2007, MNRAS, 380, 757
\bibitem[\protect\citeauthoryear{Nemmen et al.}{2012}]{nemmen2012}Nemmen R.S., Georganopoulos M., Guiriec S., Meyer E.T., Gehrels N., Sambruna R.M., 2012, Science, 338, 1445
\bibitem[\protect\citeauthoryear{Panaitescu \& Kumar}{2000}]{panaitescu2000}Panaitescu A., Kumar P., 2000, ApJ, 543, 66
\bibitem[\protect\citeauthoryear{Panaitescu \& Kumar}{2002}]{panaitescu2002}Panaitescu A., Kumar P., 2002, ApJ, 571, 779
\bibitem[\protect\citeauthoryear{Pe'er}{2012}]{pe'er2012}Pe'er, A., 2012, ApJL, 752, L8
\bibitem[\protect\citeauthoryear{Piran}{2004}]{piran2004}Piran T., 2004, RvMP, 76, 1143
\bibitem[\protect\citeauthoryear{Piro et al.}{2005}]{piro2005}Piro L. et al., 2005, ApJ, 623, 314
\bibitem[\protect\citeauthoryear{Rezzolla et al.}{2011}]{rezzolla2011}Rezzolla L., Giacomazzo B., Baiotti L., Granot J., Kouveliotou C., Aloy M.A., 2011, ApJL, 732, L6
\bibitem[\protect\citeauthoryear{Rhoads}{1997}]{rhoads1997}Rhoads J.E., 1997, ApJ, 487, L1
\bibitem[\protect\citeauthoryear{Rhoads}{1999}]{rhoads1999}Rhoads J.E., 1999, ApJ, 525, 737
\bibitem[\protect\citeauthoryear{Ruffert \& Janka}{1999}]{ruffert1999}Ruffert M., Janka H.-Th., 1999, A\&A, 344, 573
\bibitem[\protect\citeauthoryear{Sari et al.}{1998}]{sari1998}Sari R., Piran T., Narayan R., 1998, ApJ, 497, L17
\bibitem[\protect\citeauthoryear{Sari et al.}{1999}]{sari1999}Sari R., Piran T., Halpern J.P., 1999, ApJ, 519, L17
\bibitem[\protect\citeauthoryear{Scannapieco et al.}{2003}]{scannapieco2003}Scannapieco E., Schneider R., Ferrara A., 2003, ApJ, 589, 35
\bibitem[\protect\citeauthoryear{Starling et al.}{2008}]{starling2008}Starling R., van der Horst A., Rol E., Wijers R., Kouveliotou C., Wiersema K., Curran P., Weltevrede P., 2008, ApJ, 672, 433
\bibitem[\protect\citeauthoryear{Suwa \& Ioka}{2011}]{suwa2011}Suwa Y., Ioka K., 2011, ApJ, 726, 107
\bibitem[\protect\citeauthoryear{Tingay et al.}{2013}]{tingay2013}Tingay et al., 2013, PASA, 30, 7
\bibitem[\protect\citeauthoryear{Toma et al.}{2011}]{toma2011}Toma K., Sakamoto T., M\'{e}sz\'{a}ros P., 2011, ApJ, 731, 127
\bibitem[\protect\citeauthoryear{Trenti \& Stiavelli}{2009}]{trenti2009}Trenti M., Stiavelli M., 2009, ApJ, 694, 879
\bibitem[\protect\citeauthoryear{Tumlinson}{2006}]{tumlinson2006}Tumlinson J., 2006, ApJ, 641, 1
\bibitem[\protect\citeauthoryear{van der Horst et al.}{2005}]{vanderhorst2005}van der Horst A., Wijers R., Rol E., 2005, NCimC, 28, 467
\bibitem[\protect\citeauthoryear{Wijers \& Galama}{1999}]{wijers1999}Wijers R., Galama T., 1999, ApJ, 523, 177
\bibitem[\protect\citeauthoryear{Wise et al.}{2012}]{wise2012}Wise J., Turk M., Norman M., Abel T., 2012, ApJ, 745, 50
\bibitem[\protect\citeauthoryear{Woosley}{1993}]{woosley1993}Woosley S.E., 1993, ApJ, 405, 273
\bibitem[\protect\citeauthoryear{Yonetoku et al.}{2005}]{yonetoku2005}Yonetoku, D., Yamazaki, R., Nakamura, T., Murakami, T., 2005, MNRAS, 362, 1114
\bibitem[\protect\citeauthoryear{Yoon et al.}{2012}]{yoon2012}Yoon S.-C., Dierks A., Langer N., 2012, A\&A, 542, A113
\bibitem[\protect\citeauthoryear{Yost et al.}{2003}]{yost2003}Yost S., Harrison F., Sari R., Frail D., 2003, ApJ, 597, 459
\bibitem[\protect\citeauthoryear{Zhang}{2007}]{zhang2007}Zhang B., 2007, AdSpR, 40, 1186
\end{thebibliography}
\end{document}